%% file: draft.tex
\documentclass[envcountsame,runningheads,a4paper]{llncs}
\usepackage{amsmath,amssymb,listings,graphicx,url,anyfontsize}%,stmaryrd
\usepackage[varg]{txfonts}
\usepackage[all]{xy}
\newcommand{\keywords}[1]{\par\addvspace\baselineskip
\noindent\keywordname\enspace\ignorespaces#1}
\mainmatter % start of an individual contribution

\newcommand{\negop}[1]{\operatorname{\neg}{#1}}

\newcommand{\forallop}[1]{\operatorname{\forall}{#1}}

\newcommand{\pair}[2]{\langle #1,#2 \rangle}

\newcommand{\ttFetch}{\texttt{Fe}}
\newcommand{\ttIssue}{\texttt{Is}}
\newcommand{\ttExecute}{\texttt{Ex}}
\newcommand{\ttReflect}{\texttt{Re}}

\newcommand{\F}[4]{\ttFetch^{#1}_{#2} \:  #3  \: #4}
\newcommand{\I}[4]{\ttIssue^{#1}_{#2} \:  #3 \:  #4}
\newcommand{\E}[6]{\ttExecute^{#1}_{#2} \:  #3 \:  #4 \:  #5 \:  #6}
\newcommand{\R}[7]{\ttReflect^{#1}_{#2} \: {\left[#3 \!\mathbin\Rightarrow\! #4\right]} \:  #5 \:  #6 \:  #7}

\newcommand{\ttNop}{\texttt{Nop}}
\newcommand{\ttMove}{\texttt{Move}}
\newcommand{\ttLoad}{\texttt{Load}}
\newcommand{\ttStore}{\texttt{Store}}

\newcommand{\ttJump}{\texttt{Jump}}

\newcommand{\Nop}{\ttNop}
\newcommand{\Move}[2]{\ttMove \: {#1} \: {#2}}
\newcommand{\Load}[2]{\ttLoad \: {#1} \: {#2}}
\newcommand{\Store}[2]{\ttStore \: {#1} \: {#2}}
\newcommand{\JumpIf}[2]{\ttJump \: \allowbreak #1 \ \texttt{if} \ #2}

\title{Reducing State Explosion for Software Model Checking with Relaxed Memory Consistency Models}
\titlerunning{Reducing State Explosion for Software Model Checking with Relaxed MCMs}
\author{Tatsuya Abe\inst{1} \and Tomoharu Ugawa\inst{2} \and Toshiyuki Maeda\inst{1} \and Kousuke Matsumoto\inst{2}}
\institute{\email{\{abet,tosh\}@stair.center}\; STAIR Lab, Chiba Institute of Technology \and \email{\{ugawa,matsumoto\}@plas.info.kochi-tech.ac.jp}\; Kochi University of Technology}

\begin{document}

\maketitle
\begin{abstract}
Software model checking suffers from the so-called state explosion
problem, and relaxed memory consistency models even worsen this
situation. What is worse, parameterizing model checking by memory
consistency models, that is, to make the model checker as flexible as
we can supply definitions of memory consistency models as an input,
intensifies state explosion.  This paper explores specific reasons for
state explosion in model checking with multiple memory consistency
models, provides some optimizations intended to mitigate the problem,
and applies them to McSPIN, a model checker for memory consistency
models that we are developing. The effects of the optimizations and
the usefulness of McSPIN are demonstrated experimentally by verifying
copying protocols of concurrent copying garbage collection algorithms.
To the best of our knowledge, this is the first model checking of the
concurrent copying protocols under relaxed memory consistency models.
\end{abstract}

\section{Introduction}\label{sec:introduction}

Modern computing systems are based on concurrent/parallel processing
designs for their performance advantages, and programs therefore must
also be written to exploit these designs.  However, writing such
programs is quite difficult and error-prone, because humans cannot
exhaustively consider the behaviors of computers very well. One
approach to this problem is to use software model checking, in which all possible states that can be
reached during a program's execution are explored.  Many such model
checkers have been developed (e.g.,
\cite{nusmv,spin,prism,cbmc,ams2012pgas,ams2013hips}).

However, most existing model checkers adopt \emph{strict consistency}
as a Memory Consistency Model (MCM) on shared memories, which only
allows interleaving of instruction execution, and ignore
more \emph{relaxed} MCMs than strict consistency, which allow
reorderings of instructions. This is not realistic because many modern
computer architectures such as IA64, SPARC, and
POWER~\cite{ia64,sparc9,ppcarch} have adopted
relaxed MCMs.  Relaxed MCMs facilitate the performance of
parallel-processing implementations because instructions may be
reordered and multiple threads may observe distinct views on shared
memory while strict consistency, which requires synchronization at
each memory operation, is prohibitively expensive to be implemented on
computer architectures.

As interest in MCMs has grown, some model checkers have introduced
support for them~\cite{journals/sigarch/Jonsson08,conf/spin/LindenW10,conf/spin/LindenW11,Linden13}.
However, these have been specific to certain MCMs, such as Total Store
 Ordering (TSO) and Partial Store Ordering (PSO)~\cite{Adve1996}.
We are in the process of developing a model checker,
McSPIN~\cite{mcspin}, that can handle multiple
MCMs~\cite{am2013pgas,am2014hips,am2014pgas,am2016sttt}. McSPIN can take an MCM
as an input with a program to be verified. It has a specification
language that covers various MCMs including TSO, PSO, Relaxed Memory Ordering (RMO), acquire and
release consistency~\cite{cplusplus11}, Itanium
MCM~\cite{itanium}, and UPC MCM~\cite{upc}.  By using McSPIN,  we can easily model
check a \emph{fixed} program under \emph{various} MCMs.

However, software model checking suffers from the \emph{state
explosion problem}, and relaxed MCMs even worsen this, because the
reordering of instructions allowed under relaxed MCMs enormously
increases the number of reachable states.  What is worse,
parameterizing model checking by MCMs, that is, to make the model
checker as flexible as we can supply definitions of MCMs as an input
intensifies the state explosion.

This paper explains how model checking with multiple MCMs increases
the number of reachable states, and clarifies the reasons
for state explosion specific to model checking with multiple MCMs. In
addition, some optimizations are provided that reduce state explosion,
and their effects are demonstrated through experiments.  The ideas
behind the optimizations are simple: Pruning traces, partial order
reduction, and predicate abstraction are well known to reduce state
explosion in conventional model checking~\cite{Graf97a}.  In our
former paper~\cite{am2014pgas}, we arranged pruning traces and partial
order reduction for model checking with relaxed MCMs. In this paper,
we arrange predicate abstraction, and propose \emph{stages}, which are
integrations of states under relaxed MCMs.

Although the optimization in our earlier work have enabled verification
of non-toy programs such as Dekker's mutual exclusion algorithm~\cite{am2014pgas},
it was difficult to apply McSPIN to larger problems such as verifications of
copying protocols of Concurrent Copying Garbage Collection algorithms
(CCGCs), due to the state explosion.
In this paper, we demonstrated the optimizations above enables McSPIN
to verify larger programs; we checked if a desirable property of CCGCs,
``in a single thread program, what the program reads is what it has
most recently written'',
are held or not for several CCGCs on multiple MCMs.
Though we used verifications of GCs as examples in this paper,
safety of GC is an important issue in the field (e.g.,~\cite{DoligezDam1994a,GammieHE15}),
and this achievement is a positive development.  To the best of our
knowledge, this is the first model checking of copying protocols of
CCGCs with relaxed MCMs.

%% Although the optimizations in our earlier work have enabled
%% verification of non--toy programs such as Dekker's mutual exclusion
%% algorithm~\cite{am2014pgas}, those presented here allow verifications
%% of larger programs, in particular, show a property derived from the \emph{safety}, ``what the
%% mutator reads is what it has most recently written in a single thread
%% program'', of copying protocols of Concurrent Copying Garbage
%% Collection algorithms (CCGCs). While there exist several properties to
%% be verified on CCGCs (e.g., \cite{DoligezDam1994a}), we handle
%% a \emph{single} property and \emph{one} mutator since
%% this paper focuses on optimizations for model checking with multiple
%% MCMs.  Nevertheless, this achievement is a positive development in the
%% field. For example, Chicken~\cite{chicken}, a CCGC designed for
%% x86-TSO~\cite{journals/cacm/SewellSONM10}, is known to work
%% incorrectly under PSO~\cite{gcbook}. In this paper, we confirm that
%% McSPIN constructs such an execution trace that violates the safety.
%% We also conduct model checking of CCGCs, Staccato~\cite{staccato} and
%% Stopless~\cite{stopless} with relaxed MCMs.  To the best of our
%% knowledge, this is the first model checking of copying protocols of
%% CCGCs with relaxed MCMs.

The rest of this paper is organized as follows: Sect.~\ref{sec:mcspin}
describes McSPIN with exploring the reasons for state explosion
specific to model checking with MCMs, and Sect.~\ref{sec:optimization}
describes the relevant optimizations we have applied in
McSPIN. Sect.~\ref{sec:experiment} presents experimental results using
McSPIN on different CCGCs and shows the effectiveness of the
optimizations. Sect.~\ref{sec:relatedwork} discusses related work, and
the conclusions and directions for future work are presented in
Sect.~\ref{sec:conclusion}.

\section{McSPIN}\label{sec:mcspin}

We first briefly review our earlier
work~\cite{am2013pgas,am2014hips,am2014pgas} on constructing a general
model checking framework with relaxed MCMs and developing and
implementation. In the following, we do not distinguish the framework
from its implementation and refer to both as McSPIN. In McSPIN, threads on
computers with shared memory are uniformly regarded as processes that have
their own memories. Therefore, we formally call threads (in the usual
sense) \textit{processes} (in McSPIN), while we refer to them as ``threads'' when
informally explaining behavior on shared-memory systems.
\if0
Also note that this paper numbers process identifiers,
branch counters, etc., from 0 in accordance with the current implementation of
McSPIN, which is different from the notation in our earlier work
(numbering from $1$).
\fi

\subsection{Syntax}\label{sec:program}

A program is an $N$-tuple of sequences of instructions
defined as follows:
\begin{align*}
& \mbox{(Instruction)}   & i & \Coloneqq \langle L , A , \iota \rangle \, ,\\
& \mbox{(Raw Instruction)} & \iota & \Coloneqq \Move{r}{t}\ \mid \Load{r}{x} \mid \Store{x}{t} \mid \JumpIf{L}{t} \mid \Nop \, ,\\
& \mbox{(Term)}   & t & \Coloneqq v \mid r \mid t + t \mid t - t \mid \cdots \, ,\\
& \mbox{(Attributes)}  & A & \Coloneqq \{ a,\ldots, a \} \, ,
\end{align*}
where $N$ is the number of processes.
An instruction $i$ is a triple of a label, attributes, and a raw
instruction. A label $L$ designates an instruction in a program. An
attribute $a \in A$ denotes an additional label for a raw instruction,
has no effect itself, and are used to describe constraints specified
by an MCM.

Here $r$ is variable local to a process and $x,y,\ldots$\,, are shared
variables.
The raw instruction $\Move{r}{t}$
denotes the assignment of an evaluated value of a term $t$ to
a process-local variable $r$, which does not affect
other processes. The term $v$ denotes an immediate
value. The terms $t_0 + t_1, t_0 - t_1, \ldots$\,, denote standard arithmetic
expressions. $\Load{r}{x}$ represents loading $x$
from its own memory and assigning its value to $r$. $\Store{x}{t}$
denotes storing an evaluated value of $t$ to $x$ on its own memory.
$\JumpIf{L}{t}$ denotes a conditional jump to $L$ depending on the
evaluated value of $t$. Note that $t$ contains no shared values;
to jump to $L$ depending on $x$, it is necessary to perform
$\Load{r}{x}$ in advance. $\Nop$ denotes the usual no-operation.

Careful readers may wonder why no synchronization instructions such as
\emph{memory fence} and \emph{compare-and-swap} instructions appear.
In McSPIN, a memory fence is represented as a $\Nop$ with
attribute \texttt{fence}, and its effect is defined at each input MCM,
that is, multiple types of fences can be defined.  This flexibility
enables verification of a \emph {fixed} program with \emph{different}
MCMs as explained in more detail in Sec.~\ref{sec:mcm}.
Compare-and-swap (usually an instruction on a computer architecture)
is also represented by compound statements, which can be seen in
Appendix~\ref{sec:compare_and_swap}.

Programs (inputs to McSPIN) have to be written in the assembly-like
modeling language. Such low-level languages are suitable for handling
MCMs that require one to carefully take into account effects on specific computer
architectures. However, these
languages may not be practical for writing programs. McSPIN has a C-like
modeling language to facilitate programming, but this is beyond the
scope of the present paper.

\subsection{Semantics}\label{sec:semantics}

McSPIN adopts trace semantics with states. Execution traces are
sequences of \emph{operations}, defined as follows:
\begin{align*}
& \mbox{(Operation)} & o & \Coloneqq \F{j}{q}{p}{i} \mid \I{j}{q}{p}{i} \mid \E{j}{q}{p}{i}{\ell}{v} \mid \R{j}{q}{p}{p}{i}{\ell}{v} \, .
\end{align*}
One key point in handling different MCMs is to consider at most four
kinds of operations for an instruction.  For any instruction, its
fetch and issue operations are considered.  Load and store
instructions have execution operations.  Store instructions have
reflect operations.  An effect of each operation is formally defined
in our former paper~\cite{am2016sttt}.  In this paper, we roughly
explain why such operations are introduced.

Under very relaxed MCMs such as C++~\cite{cplusplus11} and UPC
MCM~\cite{upc}, distinct threads can exhibit different program
behaviors; that is, each thread has its own execution trace.
% Of course, all the execution traces must be consistent in a sense of a given MCM.
To represent these in one trace,
we add a process identifier $q$, denoting an observer process as a
subscript of an operation. In addition, McSPIN can handle programs
with loops. To distinguish multiple operations corresponding to an
instruction, an operation has a branch counter $j$ that designates the
$j$th iteration within a loop.

We explain the four kinds of operations by example.  $\F{j}{q}{p}{i}$
denotes fetching an instruction $i$ from a process $p$, which enables
the issuance of $i$. By default, this also increments the program
counter of $p$ if $i$'s raw instruction is not $\ttJump$. If so, the
program counter is not changed and will be changed when the $\ttJump$
is issued. McSPIN is equipped with a
\emph{branch prediction} mode that can be switched on or off. In branch
prediction mode, the program counter is non-deterministicly
incremented or set to $L$ when $\JumpIf{L}{t}$ is fetched.  Thus, in
order to handle branch prediction, fetch has to be distinguished from
issue.

Although branch prediction is often ignored in specifications of MCMs,
note that no branch prediction implicitly prohibits some kinds of
reorderings across conditionals.  For example, no branch prediction on
the process-model that McSPIN adopts cannot perform the
so-called \emph{out-of-thin-air} read~\cite{Manson2005} in the program
in Table 17.6 of Java language specification~\cite{JavaLangSpec},
although \emph{legal} executions under Java MCM are specified by not
using a \emph{total} order of operations on such process-model but
consistency between \emph{partial} orders of operations on threads.

%No description about branch prediction means that reordering and
%memory fence around conditionals are obscure.  By handling branch
%prediction explicitly, McSPIN promotes formal descriptions of MCMs.

\if0
For example, although McCloskey said that his algorithm
works incorrectly under RMO ~\cite{staccato}, a load instruction to be
reordered is in a conditional statement
\fi

%
Operation $\I{j}{q}{p}{i}$ denotes the issuing of an instruction
$i$ to a process $p$. Effects that complete inside the register on $p$
(not $p$'s own memory) are performed. For example, while issuing
$\Move{r}{t}$ indicates assignment of an evaluated value of $t$ to
$r$, $\Store{x}{t}$ implies evaluation of $t$ only.  In branch
prediction mode, a predicted execution trace in fetching $\ttJump$ is
checked.

Operation $\E{j}{q}{p}{i}{\ell}{v}$ denotes execution of an
instruction $i$ on a process $p$. Effects that complete inside $p$
are performed. For example, while the execution of $\Load{r}{x}$
means that $v$ is loaded from $x$ (at location $\ell$) and assigned to
$r$, $\Store{x}{t}$ represents storing an evaluated value
$v$ of $t$ to $x$ (at location $\ell$) in $p$'s own memory.
While an instruction is issued, its (intra-process) effect may not
have occur yet.  Itanium MCM allows such situation, by distinguishing
issues from executions of instructions.

Operation $\R{j}{q}{p_0}{p_1}{i}{\ell}{v}$ denotes reflects of an
instruction $i$ from process $p_0$ to $p_1$. The reflect of
$\Store{x}{t}$ means storing an evaluated value $v$ of $t$ to $x$ at $\ell$ in $p_1$'s own memory.  While a store instruction is
executed, that is, its effect is reflected to its store buffer, its
(inter-process) effect may not be reflected to other processes yet.
One reflect may be immediately passed, and another reflect may be
delayed. Moreover, processes can observe distinct views a.k.a.\ \emph{the IRIW test}~\cite{BA2008}.  Our definition covers such situations.

While the distinction enables delicate handling effects of
instructions, it intensifies state explosion since the number of
interleavings of operations increases.

To handle more relaxed MCMs, it is also necessary to distinguish
multiple operations that are generated from an instruction in a loop
statement, whereas this is unnecessary when queues can be used to
handle specific MCMs such as TSO and PSO. For example, in a code
$(\Store{x}{r_0};
\Move{r_0}{r_0+1}; \JumpIf{0}{1}) \parallel \Load{r_0}{x}$, the second
fetch of the $\ttStore$ on the former process may follow the
fetch of the $\ttLoad$ on the latter process, while the first fetch of
the $\ttStore$ on the former process may precede it.  To the best of
our knowledge, no existing method can handle such low-level jumps (across which
instructions may be reordered) in a detailed fashion, which is
necessary for verification of CCGCs.

\subsection{Formalized Memory Consistency Models}\label{sec:mcm}

MCMs are sets of constraints that control program behaviors on the very
relaxed semantics that McSPIN adopts and are formally defined as a
first-order formula as follows:
\begin{align*}
\varphi & \Coloneqq \mathrm{x_\tau} = \mathrm{x'_\tau} \mid \mathrm{x_\tau} < \mathrm{x'_\tau} \mid \negop{\varphi} \mid \varphi \supset \varphi' \mid \forallop{\mathrm{x_\tau}} \ldotp \varphi(\mathrm{x_\tau}) \, ,
\end{align*}
where $\tau$ denotes one of Variable, Location, Label, Value,
Instruction, Raw Instruction, Attribute, Branch Counter, and
Operation.
Here $\mathrm{x}_\tau$ represents metavariables in
the syntax of McSPIN. For example, $x_{\mathrm{Location}} <
x'_{\mathrm{Location}}$ can be read as $\ell < \ell'$. In addition,
$<$ with respect to $\mathrm{Operation}$ identifies the order of execution
between operations. We use standard notation such as
$\wedge$, $\vee$, and $\exists$ and assign higher precedence to $\neg$, $\wedge$,
$\vee$, and $\supset$.
%%
%We also write $x{_\tau} < x'_{\tau}$ as $(\negop{x{_\tau} = x'_{\tau}}) \wedge x{_\tau} \leq x'_{\tau}$.

Example constraints can be seen in
\cite{am2013pgas,am2014hips,am2014pgas}, and Itanium and UPC MCMs are fully formalized in their journal version\cite{am2016sttt}.
Here we focus on only two. In Sect.~\ref{sec:program}, we stated that
the effect of a memory fence can be flexibly defined by an input
MCM. A memory fence forces evaluation of
all the reflects of store
instructions that are fetched before the memory fence. This is
represented as follows:
\[
\F{j_0}{q}{p}{i_0} < \F{j_1}{q}{p}{(L_1,A_1,\Nop)} \supset \R{j_0}{q}{p}{p_0}{i_0}{\ell_0}{v_0} < \I{j_1}{q}{p}{(L_1,A_1,\Nop)}\, ,
\]
where $\texttt{fence} \in A_1$, $i_0$'s raw instruction is $\ttStore$,
and all free variables are universally quantified. Meanwhile, we can
consider another operation that forces $\ttLoad$ only:
\[
\F{j_0}{q}{p}{i_0} < \F{j_1}{q}{p}{(L_1,A_1,\Nop)} \supset \E{j_0}{q}{p}{i_0}{\ell_0}{v_0} < \I{j_1}{q}{p}{(L_1,A_1,\Nop)}\, ,
\]
where $\texttt{fence} \in A_1$ and $i_0$'s raw instruction is
$\ttLoad$.

One constraint that differentiates TSO from PSO
with \emph{multiple-copy-atomicity}~\cite{Sarkar11}, which prohibits
two threads from observing different behaviors of write operations
that the two threads do not perform, is whether reflects of store
instructions are \emph{atomically} performed \emph{in program
order}. This can be represented as follows:
\[
\F{j_0}{q}{p}{i_0} < \F{j_1}{q}{p}{i_1} \supset \R{j_0}{q}{p}{p_0}{i_0}{\ell_0}{v_0} < \R{j_1}{q}{p}{p_1}{i_1}{\ell_1}{v_1}\, ,
\]
where $i_0$'s and $i_1$'s raw instructions are $\ttStore$
instructions.  This constraint causes \emph{every} reflect of $i_1$ to
await completion of \emph{all} reflects of $i_0$.  Full constraints of
TSO, PSO, and other relaxed MCMs are formalized in McSPIN's public
repository~\cite{mcspin}.

\subsection{Translation into PROMELA}\label{sec:translation}

McSPIN uses the model checker SPIN as an engine and translates programs
written in our modeling language into PROMELA, the modeling language
of SPIN. The underlying idea is quite simple. McSPIN translates
sequential compositions of statements $i_0; i_1; \ldots$ written in
our modeling language into PROMELA loop statements as
follows:
{\small
\begin{lstlisting}
do
:: @(guard$_{0,0}$)@ -> @(operation of $\ttFetch$ of $i_0$)@; @(epilogue$_{0,0}$)@;
:: @(guard$_{0,1}$)@ -> @(operation of $\ttIssue$ of $i_0$)@; @(epilogue$_{0,1}$)@;
:: @(guard$_{0,2}$)@ -> @(operation of $\ttExecute$ of $i_0$)@; @(epilogue$_{0,2}$)@;
:: @(guard$_{0,3}$)@ -> @(operation of $\ttReflect$ of $i_0$ to $p_0$)@; @(epilogue$_{0,3}$)@;
:: ...
:: @(guard$_{0,(N-1)+3}$)@ -> @(operation of $\ttReflect$ of $i_0$ to $p_{N-1}$)@; @(epilogue$_{0,(N-1)+3}$)@;
:: @(guard$_{1,0}$)@ -> @(operation of $\ttFetch$ of $i_1$)@; @(epilogue$_{1,0}$)@;
:: ...
:: else -> break;
od;
\end{lstlisting}
}
A PROMELA loop statement has multiple clauses with guards. One of
those clauses whose guards are satisfied is non-deterministicly
chosen and processed.  Let $\texttt{clock}$ be a time counter.  Each
clause corresponds to performing an operation as follows:
{\small
\begin{lstlisting}
 end_@$o$@==0 -> @$o$@; end_@$o$@=clock; clock++;
\end{lstlisting}
}
\noindent
where the positiveness of $\texttt{end\_}o$ denotes that $o$ has already
performed.

Although such a PROMELA code may admit very relaxed behavior that does
not satisfy an input MCM, McSPIN appropriately removes such execution
traces. Assertions can be written not only at the end of
a program, but also at any place within. This is
important for CCGC verification, because we would like to confirm data
consistency at a certain place and moment. McSPIN modifies
assertion statements to follow the input MCM. Let $\varphi$ be
an assertion that we wish to verify. McSPIN adds (formalized) constraints
that an input MCM obligates to $\varphi$ as a conjunct.
For example,
the constraint that differentiates TSO and PSO, as explained in
Sect.~\ref{sec:mcm}, is translated into
{\small
\begin{lstlisting}
 !(end_{@$\F{j_0}{q}{p}{i_0}$@}<end_{@$\F{j_1}{q}{p}{i_1}$@})||end_{@$\R{j_0}{q}{p}{p_0}{i_0}{\ell_0}{v_0}$@}<end_{@$\R{j_1}{q}{p}{p_1}{i_1}{\ell_1}{v_1}$@}
\end{lstlisting}
}
\noindent
and added to the assertion $\varphi$ as a conjunct, where \texttt{!}
and \texttt{||} represent negation and disjunction in PROMELA, respectively.
Thus, execution traces that violate the MCM are removed when
assertions are checked.

\section{Optimizations}\label{sec:optimization}

Here we provide MCM-sensitive optimization techniques to reduce the
problem specific to model checking with multiple MCMs.  The
optimizations described in Sects.~\ref{sec:pruning}
and \ref{sec:predicate} were introduced in \cite{am2014pgas}; we
briefly review them here in order to make it easy to understand an
optimization introduced in Sect.~\ref{sec:stage}.

\subsection{Enhanced Guards: Pruning Inadmissible Execution Traces}\label{sec:pruning}

As explained in Sect.~\ref{sec:translation}, McSPIN explores all
execution traces and removes traces that are inadmissible under an
input MCM in checking assertions. This is obviously redundant. A
straightforward method to prune inadmissible execution traces is to
enhance guards for clauses corresponding to operations. A guard that
is uniformly generated as $\texttt{end\_}o\texttt{==0}$ from an
operation $o$ in Sect.~\ref{sec:translation} is enhanced by an input
MCM (details are provided in~\cite{am2014pgas}). 
We explain this using the constraint that differentiates TSO
and PSO, as set out in Sect.~\ref{sec:mcm}. The constraint claims
that all reflects of $i_1$ must wait for all reflects of $i_0$,
where $i_0$ precedes
 $i_1$ in program order. McSPIN adds a condition
{\small
\begin{lstlisting}
 !(end_{@$\F{j_0}{q}{p}{i_0}$@}<end_{@$\F{j_1}{q}{p}{i_1}$@})||end_{@$\R{j_0}{q}{p}{p_0}{i_0}{\ell_0}{v_0}$@}>0
\end{lstlisting}
}
\noindent
corresponding to this claim to the guard of the reflect of
$i_1$.

\subsection{Defining Predicates: Promoting Partial Order Reduction}\label{sec:predicate}

As explained in Sect.~\ref{sec:translation}, it is necessary to judge
whether an execution trace is admissible to a given MCM. This means
that it is also necessary to remember orders between operations in the
execution trace. The most straightforward method is to use a
time counter; that is, to substitute a variable
$\texttt{end\_}{o}$ (defined at each operation) with the time at which
operation $o$ was performed.
However, time counters are too concrete to reduce state explosion. For
example, consider four operations $o_0$, $o_1$, $o_2$, $o_3$
under the constraint $o_0 < o_1 \supset o_2 < o_3$.
If times are substituted for the variables $\texttt{end\_}{o_k}$ ($0 \leq k <
4$, then the number of combinations $\langle \texttt{end\_}o_0,
\texttt{end\_}o_1, \texttt{end\_}o_2, \texttt{end\_}o_3 \rangle$ is
24 (${=}4!$), which distinguishes states more concretely than the
constraint requires.

When considering the constraint rule, it suffices to remember the
order of $o_0$ and $o_1$ and of $o_2$ and $o_3$, because nothing else
is used to define the constraint. We introduce new variables
$\texttt{ord\_}o_0\texttt{\_}o_1$ and
$\texttt{ord\_}o_2\texttt{\_}o_3$, and call them \emph{defining
 predicates} of the constraint or, formally, atomic formulas consisting
of the predicate symbol $<$ (or $\leq$) between operations that occur
in the constraint. Because the defining predicates preserve the order
of times at which the operations are performed, we change
$\texttt{end\_}{o_k}$ to boolean variables that denote whether the
operation has been performed. After all the operations have been
performed (that is, $\texttt{end\_}o_k = 1 \; (0 \leq k < 4)$), the
possible states are
$\pair{\texttt{ord\_}o_0\texttt{\_}o_1}{\texttt{ord\_}o_2\texttt{\_}o_3}
= \{\pair{0}{0}, \pair{0}{1}, \pair{1}{1}\}$, of cardinality $3$.

\subsection{Stage: Abstracting Programs by MCM-Deriving Predicates}\label{sec:stage}

Predicate abstraction~\cite{Graf97a} is one promising method to
reduce state explosion in model checking. In this subsection, we show
that predicates exist that are determined by an input MCM. Such
predicates integrate states that do not have to be separated with
respect to an input MCM. Therefore, the predicate abstractions
have no omission of checking.

To handle the effects of instructions more delicately, McSPIN has at most four
kinds of operations, $\ttFetch$, $\ttIssue$, $\ttExecute$, and
$\ttReflect$ for one instruction. However, some MCMs do not require complete
distinction.
Assume that an input MCM has the 
constraint
$
\I{j}{q}{p}{i} < o \supset \E{j}{q}{p}{i}{\ell_1}{v_1} < o
$
as called \emph{integration} in \cite{am2014hips,am2014pgas}, which indicates
that no operation can interleave two operations $\I{j}{q}{p}{i}$ and
$\E{j}{q}{p}{i}{\ell_1}{v_1}$. In an earlier version, McSPIN
generated clauses that had guards waiting for
$\E{j}{q}{p}{i}{\ell_1}{v_1}$ when $\I{j}{q}{p}{i}$ was performed.
Such guards control program behaviors in accordance with an input MCM.

In this paper, we promote integration to state level rather than
execution-trace level. In earlier versions, McSPIN generated one
clause at each operation; that is, at most $3+N$, the cardinality of ${\{\ttFetch, \ttIssue, \ttExecute\} \cup \{\,\ttReflect \:
k \mid 0 \leq k < N\,\}}$, clauses at each instruction, where $N$ is
again the number of processes, and $\ttReflect \: k$ denotes a reflect to $k$. In the current version, McSPIN can
accept additional input \emph{stages} $S = \{s_0,s_1,\ldots,s_{M-1}\}$
for an input MCM. Formally, stages are partitions of
$\{\ttFetch, \ttIssue, \ttExecute\} \cup \{\,\ttReflect \: k \mid
0 \leq k < N\,\}$. We write $f_S$ for the induced mapping from the
stages. McSPIN generates $M$ clauses at each instruction $i$, where
$M$ is the number of stages of $i$ as follows:
{\small
\begin{lstlisting}
do
:: @(guards$_{0,f_S (s_0)}$)@ -> @(operation of $f_S (s_0)$ of $i_0$)@; @(epilogue$_{0,f_S (s_0)}$)@;
:: @(guards$_{0,f_S (s_1)}$)@ -> @(operation of $f_S (s_1)$ of $i_0$)@; @(epilogue$_{0,f_S (s_1)}$)@;
:: ...
:: @(guards$_{0,f_S (s_{M-1})}$)@ -> @(operation of $f_S (s_{M-1})$ of $i_0$ to $p_{N-1}$)@; @$\!$(epilogue$_{0,f_S (s_{M-1})}$)@;
:: ...
:: else -> break;
od;
\end{lstlisting}
}

This optimized translation reduces checking space and time. By loading
such a PROMELA code, SPIN remembers not unintegrated states themselves
but stages.  This implies that state-vector on SPIN is kept small.
Memory is not, therefore, consumed so much.  This optimization also
saves time to check whether clauses are executable since the number of
clauses is smaller.

Let us see example stages for TSO and PSO with neither \emph{branch
prediction} nor \emph{multiple-copy-atomicity}~\cite{Sarkar11}, which
prohibits two threads from observing different behaviors of write
operations that the two threads do not perform.  Since these MCMs
allow $\ttLoad$s to overtake (inter-process) effects of $\ttStore$s,
each member of $\{\ttReflect \: k \mid 0 \leq k < N\,\}$ has to be
separated from $\ttExecute$.  However, $\ttFetch$, $\ttIssue$, and
$\ttExecute$ do not have to be separated.  Also, $\ttReflect \: k$
does not have to be distinguished from $\ttReflect \: k' \; (k' \neq
k)$ by multiple-copy-atomicity.  We can therefore introduce the
following stages:
\[
S = \{s_0,s_1\} \qquad\qquad f_S (s) = \begin{cases}\{\ttFetch,\ttIssue,\ttExecute\} & \mbox{if } s = s_0\\\{\,\ttReflect \: k \mid 0 \leq k < N \,\} & \mbox{if } s = s_1 \enspace . \end{cases}
\]

Given a stage $S$ (and its mapping $f_S$), McSPIN
automatically returns PROMELA code in which clauses are integrated;
in particular, guards and epilogues are appropriately generated
from an input MCM.

\if0
\subsection{Directions for Improvement}\label{sec:tips}

The following are improvements that bring McSPIN closer to
practical use.

\subsubsection{Setting Supremums of Loop Iteration at Each Thread.}\label{sec:supreme}

McSPIN adopts model checking in which loop iterations are
bounded. In its current version, McSPIN permits setting a supremum
%%
of iterations at each process. This is significant for verifying GCs
consisting of two kinds of threads, collectors and mutators, because it
enables faster verification to set different supremums for collectors
and mutators.

\subsubsection{Switching Branch Prediction at Each Jump Instruction.}\label{sec:partialprediction}

Model checking with branch prediction is an advantage of McSPIN. If
all conditionals were predicted, McSPIN would have to take into
account $2^L$ combinations, where $L$ is the number of conditionals.
In the current version, McSPIN permits on--off switching for each
$\ttJump$.
\fi

\section{Experiments}\label{sec:experiment}

\begin{minipage}{.51\textwidth}
In this section, we demonstrate the effects of the optimizations introduced in
Sect.~\ref{sec:optimization}.
The figure to the right shows our experimental environment, with ample memory.
\end{minipage}
\begin{minipage}{.49\textwidth}
\hfill{
{\small
\begin{tabular}{|ll|}
\hline
CPU: & Intel Xeon E5-2670 2.6GHz\\% $16 \mbox{cores} \times 2$\\
Memory: & DDR3-1066 1.5TB\\
SPIN: & 6.4.5 \\
GCC: & 5.3.0\\\hline
\end{tabular}
}
}
\end{minipage}
%\vspace{\baselineskip}
The optimizations described in Sects.~\ref{sec:pruning} and \ref{sec:predicate}
have enabled verification of relatively large programs such as Dekker's
algorithm~\cite{am2014pgas}.
Here we demonstrate that the optimization described in
Sect.~\ref{sec:stage} enables verification of genuinely large programs.

\subsection{Experimental Setting}\label{sec:setting}

We chose CCGCs as examples of large programs.
%Verification of
%CCGCs are not only examples, but also serious problems in
%practice~\cite{}. => Related work?
In this subsection, we briefly explain the CCGCs we used.

Garbage collection (GC) is a basic service of modern programming languages.
Its role is to find garbage, that is,
data objects that are no longer in use by the application,
and to reclaim the memory that those objects occupy.
\iffalse
To determine such \textit{garbage}
objects, GCs use an approximation: objects that are reachable
by tracing a chain of pointers from the GC \textit{roots} such as
variables of the application are \textit{live}, i.e., they may be
used in the future, and those objects that are not live are
garbage. Copying GC traverses the object graph and copies
objects while copying visited objects to a separate \textit{to-space},
so that it constructs the same topology in to-space.
\else
Copying GC accomplishes this by copying live objects, i.e.,
those that may be used in the future, to a separate space and
then releasing the old space that contains the copied objects and
garbage.
\fi
Concurrent GC, as the name suggests, runs concurrently with
the application.  What is difficult in designing CCGC algorithms is
that the garbage collector thread and an application thread may
race; the application thread may change the contents
of an object that is being copied by the garbage collector.
This may be the case even with an single thread application.
Because an application thread changes, or mutates, the object,
we call it a \textit{mutator}.
If a mutator writes to the object that is being copied, the collector
may copy a stale value, which means that the latest value gets lost.
Various copying protocols have been proposed to provide application
programmers with reasonable MCMs, all
of which require the mutators to do some work on every
read (\textit{read barrier}) or write (\textit{write barrier})
operation or both, in which the mutator synchronizes with the collector.

Because such barriers incur overhead for every read or write operation,
one goal of CCGC algorithms is to design barriers that are as lightweight
as possible.
Thus, synchronizations such as compare-and-swap should be minimized.
With relaxed MCMs, memory barriers should also be minimized.
Unfortunately, the synchronizations required for safety depend
on the given MCM; it is often the case that those synchronizations that are
redundant for one MCM are mandatory for another.

\subsubsection{Model.}

% The property
We experimentally checked the safety of concurrent
copying protocols, \textit{in a single thread program, what the
mutator reads is what it has most recently written}.  This property
is expected to be held in any reasonable MCMs such as the happens-before
consistency of Java~\cite{JavaLangSpec}.
%The CCGC algorithms we checked are Chicken~\cite{chicken},
%Staccato~\cite{staccato} and Stopless~\cite{stopless}.
%Brief description of these algorithms will be shown later.
The complete McSPIN models for checking this property can be
found in Appendix or the McSPIN public repository~\cite{mcspin}.
Here, we briefly explain the model.

% The common setting -- single mutator, single object, behaviours of threads
In our model checking, we made some assumptions.
We assume that there is a single mutator thread, i.e., the application
is a single thread program.  Remark that even if there is a single
mutator thread, there is another thread, the collector thread, and
they may race.
We also assume that there is only a single object with a single
integer slot in the heap.

The mutator has a pointer to the object and repeatedly
reads from and write to the object through the pointer.
\iftrue
On write operations, it remembers the value it wrote.% in a local variable.
On read operations, it checks if the read value is equal to the value
it lastly wrote.
Meanwhile, the collector copies the object following to the copying
protocol of each algorithm.  Once it successfully copied, %it \textit{flips}
%% the mutator's variable so that the variable refers to the copy of the
%% object.
the collector rewrites the mutator's pointer to the object so that the
pointer points at the copy.

\fi
% Mutator's operation -- barriers, handshake
To cooperate with the collector, the mutator uses the
read and write barriers required by the copying protocol
on its read and write operations.
For some algorithms, the mutator also performs so called the
checkpoint operation between object accessing operations,
where the mutator polls and answers collector's requests.
Some collectors request the mutator to answer the handshake
by setting a per-mutator handshake request flag.  The
checkpoint operation clears the flag to let the collector know
the mutator has observed the flag set.  In TSO, if a mutator
observes the flag is set, all stores preceding the store setting
the flag are guaranteed to be visible to the mutator.

% McSPIN Model
We created McSPIN models for each CCGC algorithms we describe below.
In the models,
the mutator has an infinite loop, where it reads or writes once
per an iteration.  It also performs a checkpoint operation
before and after each read or write.  Thus, the supremum of loop
iterations on the mutator limits the number of mutator's memory accesses.

\subsubsection{GC Algorithm}
%Here we experimentally verified the copying protocols of
%three CCGC algorithms, Chicken, Staccato, and Stopless.
%Below, we also demonstrate verifications of variants of Stopless.
%All algorithms install a \textit{forwarding pointer} in the header
%of every object so that the mutator can access the latest copy of
%the object. Every copy of an object has a forwarding pointer
%to the latest copy. 

In this paper, we checked three GC algorithms:
Chicken~\cite{chicken},
Staccato~\cite{staccato} and Stopless~\cite{stopless}.
The details of these algorithms can be found in their papers.
Here, we briefly explain their features.

Chicken and Staccato were basically the same algorithm
though they are developed independently. The only difference is
their target MCMs; Chicken is designed for the MCMs of Intel CPU such as IA64~\cite{ia64},
while Staccato's main target seems to be POWER MCM~\cite{ppcarch}.
These algorithms use compare-and-swap operations to resolve
races between the collector and a mutator.
In the IA64 MCM, the compare-and-swap is usually realized by
the instruction sequence \texttt{lock} \texttt{cmpxchg}.
This sequence implies memory fences.  As for POWER MCM,
the manual~\cite{ppcarch} shows a sample implementation of
the compare-and-swap operation that does not imply memory fences.
%In our experimentation, we modeled the compare-and-swap
%instruction with the McSPIN's construct to create an atomic block.
%Our model of the compare-and-swap instruction implies
%memory fences for TSO and does not for POWER MCM.

Stopless is a different algorithm from those two.  It uses
compare-and-swap operations that implies memory fence excessively,
hence chances of reordering are fewer.

\subsection{Effect of Optimization}
\input{stage.tbl}

In this subsection, we reveal the effectness of the \emph{stage}
optimization described in Sect.~\ref{sec:stage}.  We verified the models created
in Sect.~\ref{sec:setting} by using McSPIN with and without the
optimization.  We fixed the supremum of iteration on the collector to 1
and varied that on the mutator from 1 to 2.

Tab.~\ref{tbl:tsopso} shows the results of the verification.  Note that
any PROMELA code produced by McSPIN consumed around 170 MB of memory as
constant overhead.
As Tab.~\ref{tbl:tsopso} shows, the amounts of 
memory consumed and elapsed times are greatly reduced in all algorithms
compared with those without the optimization.  In particular,
when the supremum of loop iterations on the mutator was set to 2,
McSPIN without the optimization often required around 1 TB of memory,
which is far from reasonable.  However, a single iteration could
not detect any error even for the algorithms that actually work
incorrectly with PSO, i.e., Chicken and Stopless.
%Although, in the cases where McSPIN detected an error, it could
%get the result before consuming lots of memory by chance as the
%case of Stopless with PSO, this fact cannot ease the memory requirement.

Tab.~\ref{tbl:tsopso} also suggests that the more instructions the model had,
the more effective the stage optimization was.  For example, in Chicken
and Staccato with a single iteration, the net memory consumption was reduced
to 3.9--8.8 \%, while, in Stopless, it was reduced to 3.3 and 4.4 \%.
This is because the stage optimization reduced the number of units that
are subject to reordering, or clauses of the do-loop in the PROMELA code.

%%
%\input{predict.tbl}

\iffalse
Chicken is designed for x86-TSO and known to work incorrectly under
PSO~\cite{gcbook}. McSPIN can demonstrate this by constructing an
execution trace that violates an assertion using a reasonable memory
consumption that does not necessarily need a huge computing system.
%Because the execution trace does not occur when the supremum of
%iterations is set to 1, it is significant to enable model checking
%with multiple iterations.

Staccato is designed for POWER MCM.  Thus, it worked correctly
with a stronger MCM of PSO.

Stopless is also designed for x86-TSO.  Although we expected that it
works correctly even with PSO since it uses a compare-and-swap
excessively, McSPIN with stages revealed that it is wrong with
only a few megabytes of memory.
\fi

To the best of our knowledge, this is the first model checking of
these algorithms with PSO, due to the optimizations given in this paper.

\subsection{Reducing Memory Fences of Staccato}\label{sec:staccato_fence}

\input{staccato.tbl}

Because Staccato is designed for the relaxed MCM of POWER,
some memory fences are redundant on a stricter MCM. Thus, we designed and
verified a variant of Staccato for a PSO MCM with a compare-and-swap
that does not imply memory fences.  In addition, we created
an incorrect variant that lacks mandatory fences for the PSO MCM.
These variants are labeled \texttt{staccato\_pso} and \texttt{staccato\_bug}.

The result of verification is shown in Tab.~\ref{tbl:staccatovariants}.
The verification is conducted with the stage optimization.
The result of \texttt{staccato\_bug} shows that McSPIN detected an
error if we reduced fences too much.

The variants of Staccato demonstrate the usefulness of McSPIN.
When we modify a GC algorithm for a machine
with some MCM that is different from the one that the GC is originally
designed for, we add or remove some synchronizations. However,
the modified model often lacks synchronizations.
McSPIN can detect such errors in the variant with a reasonable
memory consumption. This enables us to check the GC when
we are performing modifications.

\subsection{McSPIN vs.\ Hand-Coding}\label{sec:specific}

In this subsection, we compare PROMELA codes generated by McSPIN with
codes written by hand and confirm how close McSPIN is to an ideal
implementation.

\begin{figure}[t]
\begin{minipage}{.34\textwidth}
{\scriptsize%small
\begin{lstlisting}
active proctype main() {
 run mem();
 run proc0();
 ...
}
proctype proc0() {
 WRITE(x0,0);
 WRITE(x1,1);
}
...
\end{lstlisting}
}
\end{minipage}
\qquad
\begin{minipage}{.58\textwidth}
{\scriptsize%small
  \begin{lstlisting}
proctype mem() {
endmem:
 do
 ::atomic{COMMIT_WRITE(queue_proc0);}
 ::atomic{COMMIT_WRITE(queue_proc1);}
 ...
 od;
}
inline WRITE (var,val) {...}
inline COMMIT_WRITE (queue) {...}
\end{lstlisting}
}
\end{minipage}
\vspace{-\baselineskip}
\caption{Hand-written code}\label{fig:genmmsnip}
\vspace{-\baselineskip}
\end{figure}

Whereas McSPIN generates uniform PROMELA codes that contain variables
to remember orders between operations, etc., to support different
MCMs, some variables are essentially unnecessary for verifications
specific to TSO and PSO. Because TSO never reorders store
instructions, queues (for all shared variables) at each thread to
buffer effects of write instructions suffice for verifications under
TSO as shown in Fig.~\ref{fig:genmmsnip}.
The two \texttt{WRITE}s put $\pair{\texttt{x0}}{\texttt{0}}$ and
$\pair{\texttt{x1}}{\texttt{1}}$ into the queue in order. Reflects
from the queue to shared memory are performed by
\texttt{COMMIT\_WRITE}s on a process \texttt{mem}. We omit the implementation details.
For PSO, one queue at \emph{each} shared variable is enough to
reorder the effects of write instructions to distinct shared variables.

\begin{table}[t]
\scriptsize
 \caption{Comparison between McSPIN and hand-coding}\label{tbl:genmm}
%\vspace{-\baselineskip}
\centering
\begin{tabular}{|l|r|r|r|r|r|r|r|r|r|r|r|r|r|r|} \hline
 & \multicolumn{6}{c|}{TSO} & \multicolumn{6}{c|}{PSO} \\ \cline{2-13}
 & \multicolumn{3}{c|}{McSPIN} & \multicolumn{3}{c|}{hand-written} & \multicolumn{3}{c|}{McSPIN} & \multicolumn{3}{c|}{hand-written} \\ \hline
 & state & memory & time & state & memory & time & state & memory & time & state & memory & time\\
 &       & (MB) & (sec.) &       & (MB) & (sec.) &       & (MB) & (sec.) &       & (MB) & (sec.) \\ \hline
1 & 25& 0.006& 0.01& 19& 0.002& 0.01& 25& 0.006& 0.02& 19& 0.003& 0.01\\\hline
2 & 52& 0.017& 0.02& 60& 0.009& 0.00& 65& 0.021& 0.02& 79& 0.017& 0.01\\\hline
3 & 116& 0.053& 0.02& 149& 0.026& 0.00& 241& 0.110& 0.05& 337& 0.095& 0.01\\\hline
4 & 241& 0.153& 0.04& 313& 0.064& 0.02& 977& 0.619& 0.20& 1,405& 0.504& 0.02\\\hline
5 & 457& 0.391& 0.08& 585& 0.143& 0.01& 3,985& 3.405& 0.98& 5,749& 2.500& 0.07\\\hline
6 & 800& 0.897& 0.18& 1,004& 0.276& 0.01& 16,145& 18.107& 6.13& 23,269& 11.894& 0.31\\\hline
7 & 1,312& 1.882& 0.35& 1,615& 0.493& 0.01& 65,041& 93.290& 34.06& 93,637& 54.294& 1.66\\\hline
8 & 2,041& 3.659& 0.61& 2,469& 0.829& 0.02& 261,137& 468.195& 171.10& 375,685& 246.497& 8.28\\\hline
\end{tabular}
%\vspace{-\baselineskip}
%\end{center}
%}
\end{table}

Tab.~\ref{tbl:genmm} compares PROMELA codes generated by McSPIN with
those written by hand where the constant overhead is removed. The programs are simple, consisting of multiple store
instructions (without loops). Verified properties are fixed to be
true. Each column is similar to Tab.~\ref{tbl:tsopso}.
%with the numbers of
%instructions and iterations and the types of property checking
%omitted.
The digits in the names of the codes denote the number
of store instructions at each thread, respectively.
The number of states almost coincides. Slight differences appear to
derive from the current implementation of SPIN, because we observe
that SPIN returns fewer states for a PROMELA code with a loop statement
and control variables (such as code generated by McSPIN)
than another PROMELA code with a sequential composition of statements
(like hand-written code). However, we have not investigated this in
detail.

McSPIN consumes more memory and time. This is a result of the sizes
of the state vectors and is inevitable, because McSPIN defines more
variables to determine program structures than hand-written codes, as
explained in the beginning of this subsection.

\section{Related Work}\label{sec:relatedwork}

There exists no work, which is directly compared with our work, of
model checking to take multiple MCMs in a uniform way.  Therefore, we
can find no work for its optimization has been studied.

Jonsson's seminal work discovered the potential of SPIN for program
translation toward model checking with relaxed
MCMs~\cite{journals/sigarch/Jonsson08}. However, he could not conduct
a large number of experiments, because his program translation was not
completely automatic and optimized. This paper has addressed the
problems that he left open. McSPIN supports various MCMs and takes an MCM
as an input, and its program translation is automatic. McSPIN is
greatly optimized and enables verification of larger concurrent
algorithms such as copying protocols of CCGCs.

Linden et al.~\cite{conf/spin/LindenW10,conf/spin/LindenW11,Linden13}
tackled the state explosion problem by representing store buffers as
automata. However, they handled relatively strict relaxed MCMs such as
TSO and PSO, unlike McSPIN. It is an open issue to extend their
representation so as to handle more relaxed MCMs and apply it to
McSPIN.

Modex~\cite{modex}, a model extractor of SPIN that is guided by a
user-defined test-harness, translates C codes into PROMELA codes.
However, Modex ignores relaxed MCMs.  Although revising Modex so as to
handle relaxed MCMs is surely one approach, we have developed McSPIN in
order to show the potential of program translation toward model
checking with relaxed MCMs with no restriction derived from the existing tool.

Travkin et al.~\cite{conf/hvc/TravkinMW13} developed a similar tool that translates programs into
PROMELA codes and uses SPIN as the engine for model checking,
demonstrated verifications of linearizability of concurrent algorithms
under TSO, and planned to tackle PSO.
However, their translator, which generates codes
that are similar to hand-written PROMELA code as introduced in
Sect.~\ref{sec:specific}, cannot be immediately applied to relaxed
MCMs beyond PSO. Unlike their approach, ours supports relaxed
MCMs by virtue of constructing a base that allows such relaxed behaviors and then
defines MCMs as constraints on the base. Although an issue of our
approach is addressing the state explosion problem, this paper has
presented optimizations for the problem.

Dan et al.~\cite{Dan2013} reported high utility of predicate
abstractions in model checking with relaxed MCMs by
verifying some programs with predicate abstractions under TSO and
PSO. They proposed the notion of predicate
\emph{extrapolation} to abstract a boolean program for an input
program. Although the stages introduced in this paper can be regarded as
predicate abstractions, there is a difference in usage: McSPIN
considers at most four kinds of multiple states at one instruction to support various MCMs beyond TSO/PSO. Although it is
necessary to handle the worst case under the most relaxed MCM, this is
not always the case. Stages are states that are integrated by
predicates that are uniformly generated by an input MCM.
Therefore, abstractions by the predicates never leak out of checking.
Dan et al.'s technique of extrapolating predicates seems to
be compatible with stages, and its combination with stages is an open
issue.

Theorem proving in program logic is also one promising approach to
program verification with relaxed
MCMs~\cite{DBLP:conf/vstte/Ridge10,Ferreira2010,vn13,am2016pro,am2016aplas}.
Formal verifications of GC algorithms with relaxed MCMs using theorem
provers have recently appeared~\cite{GammieHE15}. However, fully
automated verification by model checking is usually preferable to
manual (or semi-automatic) construction of proofs in theorem proving.

\section{Conclusion and Future Work}\label{sec:conclusion}

We have explained the reasons for the state explosion problem specific
to model checking with multiple MCMs, presented optimizations modified
from pruning execution traces, partial order reduction, and predicate
abstraction, and applied them to McSPIN, our model checker with
MCMs. We have also shown the effectiveness of the optimizations
through experiments of verifications of copying protocols of CCGCs,
which are larger programs.

There are four future directions for this work.  Although we verified copying protocols of CCGCs as examples of large
programs in this paper, a verification of GC algorithm is itself
subject of our interest.  Our future work includes verifications of
wide range of GC algorithms and other properties such as wait-freedom
for Chicken, which the authors designed as a wait-free
CCGC~\cite{chicken}.  These verifications may require more complicated
settings including pointers and/or multiple mutators, which need still
larger models.
The second is to
show a verification of concurrent
copying protocols with MCMs that are more relaxed than PSO.  An
advantage of McSPIN is its ability to support various MCMs.
The third is to show more realistic benchmark programs, e.g., SV-COMP benchmarks~\cite{svcomp}.
The fourth is further optimization of McSPIN to verify even larger programs.
%is also %a topic of future work.

%Further
%verification of program behaviors of concurrent copying protocols will
%be a topic of future work.  Also, it is another topic of future work
%to enhance McSPIN to handle larger programs, e.g., other kinds of GCs.

\paragraph{Acknowledgments}
The authors thank the anonymous reviewers for several comments to
improve the final version of the paper.
%In this research work we used the supercomputer of ACCMS, Kyoto University. % format by Kyoto Univ. rule
This research partly used computational resources under Collaborative Research Program for Young Scientists provided by Academic Center for Computing and Media Studies, Kyoto University. % format by Kyoto Univ. rule
This work was supported by JSPS KAKENHI Grant Numbers 25871113, 25330080, and 16K21335.

\bibliographystyle{splncs_srt} \bibliography{draft}
\input{appendix}

\end{document}

%% file: stage.tbl
\begin{table}[t]
\caption{Effects of optimization: In TSO, a compare-and-swap instruction
implies memory fences. In PSO, it does not.
Columns labeled with ``col.'' and ``mut.'' list
the number of instructions of the collector and the mutator, respectively.
Column labeled with ``loop'' lists the supremum of loop iterations on
the mutator.  For verification either with or without stages, the
first column shows the results; $\checkmark$ means no error was found
and $\times$ means a violation was found.
%the correctness property was violated.
The following columns are the number of
state transitions, the amount of memory consumed, and elapsed
times for verification, respectively.  Columns labeled with ``mem. ratio''
and ``time ratio'' list the ratios of memory and time consumption
for verification with/without stages.  The column labeled with
``net mem. ratio'' lists those that do not count constant overhead.
}\label{tbl:tsopso}
\scriptsize
\centering
\begin{tabular}{|c|l|r|r|r|r|r|r|r|r|r|r|r|r|r|r|} \hline
& & & & & \multicolumn{4}{c|}{without stages} & \multicolumn{4}{c|}{with stages} & mem. & net mem. & time \\ \cline{6-13}
MCM & algorithm & col. & mut. & loop    & & state & memory & time & & state & memory & time & ratio & ratio & ratio\\
& & & &    & & (K) & (MB) & (sec.) & & (K) & (MB) & (sec.) & (\%) & (\%) & (\%) \\\hline\hline

\input{stage.dat}
\end{tabular}
\end{table}

%% file: stage.dat
     & \raisebox{-5pt}[0ex][0ex]{chicken}   &   \raisebox{-5pt}[0ex][0ex]{24}  &  \raisebox{-5pt}[0ex][0ex]{42}  &  1  &  $\checkmark$  & 108 & 8,595 & 132 &  $\checkmark$  & 23 & 908 & 8 & 10.6 & 8.8 & 6.2\\\cline{5-16}
                                                                                                     &   &   &   &  2  &  $\checkmark$  & 2,506 & 546,038 & 8,637 &  $\checkmark$  & 534 & 24,960 & 433 & 4.6 & 4.5 & 5.0\\\cline{2-16}
TSO  &  \raisebox{-5pt}[0ex][0ex]{staccato}  &  \raisebox{-5pt}[0ex][0ex]{32}  &  \raisebox{-5pt}[0ex][0ex]{46}  &  1  &  $\checkmark$  & 141 & 14,918 & 236 &  $\checkmark$  & 26 & 1,032 & 11 & 6.9 & 5.8 & 4.7\\\cline{5-16}
                                                                                                     &   &   &   &  2  &  $\checkmark$  & 3,888 & 1,097,491 & 16,022 &  $\checkmark$  & 733 & 43,432 & 735 & 4.0 & 3.9 & 4.6\\\cline{2-16}
     &  \raisebox{-5pt}[0ex][0ex]{stopless}  &  \raisebox{-5pt}[0ex][0ex]{33}  &  \raisebox{-5pt}[0ex][0ex]{87}  &  1  &  $\checkmark$  & 90 & 28,183 & 378 &  $\checkmark$  & 14 & 1,404 & 19 & 5.0 & 4.4 & 5.0\\\cline{5-16}
                                                                                                     &   &   &   &  2  &  $\checkmark$  & 564 & 564,635 & 7,705 &  $\checkmark$  & 87 & 18,885 & 585 & 3.3 & 3.3 & 7.6\\\hline
     &  \raisebox{-5pt}[0ex][0ex]{chicken}   &  \raisebox{-5pt}[0ex][0ex]{24}  &  \raisebox{-5pt}[0ex][0ex]{42}  &  1  &  $\checkmark$  & 308 & 25,208 & 430 &  $\checkmark$  & 65 & 1,652 & 28 & 6.6 & 5.9 & 6.6\\\cline{5-16}
                                                                                                     &   &   &   &  2  &  $\times$  & 1,136 & 264,857 & 4,248 &  $\times$  & 237 & 12,190 & 227 & 4.6 & 4.5 & 5.3\\\cline{2-16}
PSO  &  \raisebox{-5pt}[0ex][0ex]{staccato}  &  \raisebox{-5pt}[0ex][0ex]{32}  &  \raisebox{-5pt}[0ex][0ex]{46}  &  1  &  $\checkmark$  & 143 & 15,166 & 243 &  $\checkmark$  & 26 & 1,032 & 11 & 6.8 & 5.7 & 4.7\\\cline{5-16}
                                                                                                     &   &   &   &  2  &  $\checkmark$  & 4,020 & 1,144,602 & 16,975 &  $\checkmark$  & 751 & 44,920 & 768 & 3.9 & 3.9 & 4.5\\\cline{2-16}
     &  \raisebox{-5pt}[0ex][0ex]{stopless}  &  \raisebox{-5pt}[0ex][0ex]{33}  &  \raisebox{-5pt}[0ex][0ex]{87}  &  1  &  $\checkmark$  & 177 & 55,210 & 833 &  $\checkmark$  & 30 & 2,520 & 46 & 4.6 & 4.3 & 5.5\\\cline{5-16}
                                                                                                     &   &   &   &  2  &  $\times$  & 45 & 45,416 & 630 &  $\times$  & 8 & 2,148 & 41 & 4.7 & 4.4 & 6.6\\\hline

%% file: staccato.tbl
\begin{table}[t]
\scriptsize
\caption{Variants of Staccato}\label{tbl:staccatovariants}
\centering
\begin{tabular}{|l|r|r|r|r|r|r|r|r|r|r|r|} \hline
 & & & & \multicolumn{4}{c|}{TSO} & \multicolumn{4}{c|}{PSO} \\ \cline{5-12}
algorithm & col. & mut. & loop    & & state & memory & time & & state & memory & time\\
 & & &    & & (K) & (MB) & (sec.) & & (K) & (MB) & (sec.) \\ \hline
%%%%\raisebox{-5pt}[0ex][0ex]{staccato} & \raisebox{-5pt}[0ex][0ex]{32} & \raisebox{-5pt}[0ex][0ex]{46}      & 1 & $\checkmark$ & 25675& 1032.083& 11.11 &  $\checkmark$ & 25871& 1032.083& 11.47\\\cline{4-12}
%%%%                                                                                                     & & & 2 & $\checkmark$ & 733271& 43432.444& 734.64 & $\checkmark$ & 751323& 44920.175& 768.46\\\hline
%%%%\raisebox{-5pt}[0ex][0ex]{staccato\_pso} & \raisebox{-5pt}[0ex][0ex]{31} & \raisebox{-5pt}[0ex][0ex]{44} & 1 & $\checkmark$ & 24574& 908.106& 9.90& $\checkmark$ & 25190& 1032.083& 10.24\\\cline{4-12}
%%%%                                                                                                     & & & 2 & $\checkmark$ & 719298& 38721.292& 637.38& $\checkmark$ & 754590& 40952.890& 703.21\\\hline
%%%%\raisebox{-5pt}[0ex][0ex]{staccato\_bug} & \raisebox{-5pt}[0ex][0ex]{30} & \raisebox{-5pt}[0ex][0ex]{44} & 1 & $\checkmark$ & 27922& 1032.083& 10.80& $\checkmark$ & 34850& 1156.061& 13.59\\\cline{4-12}
%%%%                                                                                                     & & & 2 & $\checkmark$ & 819314& 43184.488& 726.17& $\times$ & 235229& 12809.961& 216.64\\\hline
\input{staccato.dat}
\end{tabular}
\end{table}

%% file: staccato.dat
\raisebox{-5pt}[0ex][0ex]{staccato}  &  \raisebox{-5pt}[0ex][0ex]{32}  &  \raisebox{-5pt}[0ex][0ex]{46}       &  1  &  $\checkmark$  & 26 & 1,032 & 11 &   $\checkmark$  & 26 & 1,032 & 11\\\cline{4-12}
                                                                                                      &   &   &  2  &  $\checkmark$  & 733 & 43,432 & 735 &  $\checkmark$  & 751 & 44,920 & 768\\\hline
\raisebox{-5pt}[0ex][0ex]{staccato\_pso}  &  \raisebox{-5pt}[0ex][0ex]{31}  &  \raisebox{-5pt}[0ex][0ex]{44}  &  1  &  $\checkmark$  & 25 & 908 & 10 &  $\checkmark$  & 25 & 1,032 & 10\\\cline{4-12}
                                                                                                      &   &   &  2  &  $\checkmark$  & 719 & 38,721 & 637 &  $\checkmark$  & 755 & 40,953 & 703\\\hline
\raisebox{-5pt}[0ex][0ex]{staccato\_bug}  &  \raisebox{-5pt}[0ex][0ex]{30}  &  \raisebox{-5pt}[0ex][0ex]{44}  &  1  &  $\checkmark$  & 28 & 1,032 & 11 &  $\checkmark$  & 35 & 1,156 & 14\\\cline{4-12}
                                                                                                      &   &   &  2  &  $\checkmark$  & 819 & 43,184 & 726 &  $\times$  & 235 & 12,810 & 217\\\hline

%% file: appendix.tex
\appendix

\lstset{language=C,
  basicstyle=\scriptsize\ttfamily,
  columns=[l]{fullflexible},
%  numbers=left,1
  frame=single,
  numberstyle=\scriptsize,
  stepnumber=1,
  numbersep=5pt,
  escapechar=\@
  lineskip=-0.5zw
}

\section{McSPIN Model -- Compare-and-Swap}\label{sec:compare_and_swap}

Here we show the McSPIN model of compare-and-swap instructions, which we used in
the models of each copying protocols of CCGCs.  We made four variants;
\texttt{CAS} compare-and-swaps a single variable, and \texttt{CAS2} does
two variables at the same time, and those that do not report if the operation
succeeded or not, whose names have the \texttt{\_NORET} suffix.
\texttt{CAS2} is used for modeling a double-word compare-and-swap and bit fields.
The fences (\texttt{McSPIN\_fence}) appear for TSO but not for PSO.

\begin{lstlisting}
/* compare-and-swap */
static inline CAS(int target, int old, int new, int retval) {
#pragma McSPIN attribute atomic
  {
    if (target == old) {
      target = new;
      retval = 1;
    } else
      retval = 0;
  }
}

/* two-variable compare-and-swap */
static inline CAS2(int target1, int target2, int old1, int old2, int new1, int new2,
       int retval) {
#pragma McSPIN attribute atomic
  {
    if (target1 == old1)
      if (target2 == old2) {
	target1 = new1;
	target2 = new2;
	retval = 1;
      } else
	retval = 0;
    else
      retval = 0;
  }
}

/* compare-and-swap */
static inline CAS_NORET(int target, int old, int new) {
#pragma McSPIN attribute atomic
  {
    if (target == old)
      target = new;
  }
}

/* two-variable compare-and-swap */
static inline CAS2_NORET(int target1, int target2, int old1, int old2, int new1, int new2) {
#pragma McSPIN attribute atomic
  {
    if (target1 == old1)
      if (target2 == old2) {
    	target1 = new1;
    	target2 = new2;
      }
  }
}
\end{lstlisting}

\section{McSPIN Model for Chicken}

Here we show our McSPIN model for the copying protocol of Chicken.

\begin{lstlisting}
#include "stdbool.h"
#include "atomic.h"

#define INITIAL_VALUE 0

#define FROM_OBJECT 0
#define TO_OBJECT 1

int main()
{
  /* handshake */
  int hs_req;
  /* heap */
  int from_header_fwd, from_header_copying, from_body;
  int to_body;
#define to_header_fwd TO_OBJECT
#define to_header_copying FALSE
  /* mutator's variable */
  int root;

#pragma McSPIN parallel sections
  {
    /* collector */
#pragma McSPIN section
    {
      int success = 0;
      while (success == 0) {
	from_header_copying = TRUE;

	start_handshake();
	wait_for_handshake();

	/* assume: to_header_copying = FALSE */
	/* assume: to_header_fwd = TO_OBJECT */
	to_body = from_body;

	CAS2(from_header_fwd, from_header_copying, FROM_OBJECT, TRUE,  TO_OBJECT, FALSE,
             success);
      }
      root = from_header_fwd; /* flip */
    }

    /* mutator */
#pragma McSPIN section
    {
      int last_written = INITIAL_VALUE;
      int readval;
      while (true) {
#pragma McSPIN nondeterministic
	{
	  ack_handshake();
	}
#pragma McSPIN nondeterministic
	{
	  last_written = 1 - last_written;
	  write(root, last_written);
	} else {
	  read(root, readval);
	  McSPIN_assert(McSPIN_variable(last_written,1,0) == McSPIN_variable(readval,1,0));
	}
#pragma McSPIN nondeterministic
	{
	  ack_handshake();
	}
      }
    }
  }
}

static inline ack_handshake()
{
  if (hs_req == true)
    hs_req = false;
}

static inline start_handshake()
{
  hs_req = true;
}

static inline wait_for_handshake()
{
  while (hs_req != 0)
    ;
}

/* read barrier */
static inline read(int obj, int retval) {
  if (obj == FROM_OBJECT) {
    if (from_header_fwd == FROM_OBJECT)
      retval = from_body;
    else
      retval = to_body;
  } else {
    /* assume: to_header_fwd = TO_OBJECT */
    retval = to_body;
  }
}

/* write barrier */
static inline write(int obj, int val) {
  if (obj == FROM_OBJECT) {
    if (from_header_copying == TRUE) {
      CAS2_NORET(from_header_fwd, from_header_copying,	 FROM_OBJECT,     TRUE,		 FROM_OBJECT,     FALSE);
    }
    if (from_header_fwd == FROM_OBJECT)
      from_body = val;
    else
      to_body = val;
  } else {
    /* assume: to_header_copying = FALSE */
    /* assume: to_header_fwd = TO_OBJECT */
    to_body = val;
  }
}
\end{lstlisting}

\section{McSPIN Model for Staccato}

Here we show our McSPIN model for the copying protocol of Staccato.
Two fences are removed for \texttt{staccato\_pso} and another
fence is omitted for \texttt{staccato\_bug}.  They are marked with comments
in the following model.

\begin{lstlisting}
#include "stdbool.h"
#include "atomic.h"

#define INITIAL_VALUE 0

#define FROM_OBJECT 0
#define TO_OBJECT 1

int main()
{
  /* handshake */
  int hs_req;
  /* heap */
  int from_header_fwd, from_header_copying, from_body;
  int to_body;
#define to_header_fwd TO_OBJECT
#define to_header_copying FALSE
  /* mutator's variable */
  int root;

#pragma McSPIN parallel sections
  {
    /* collector */
#pragma McSPIN section
    {
      int success = 0;
      while (success == 0) {
	from_header_copying = TRUE;
	McSPIN_fence();

	start_handshake();
	wait_for_handshake();
	McSPIN_fence();       /* removed for staccato_pso and saccato_bug */
        
	/* assume: to_header_copying = FALSE */
	/* assume: to_header_fwd = TO_OBJECT */
	to_body = from_body;

	McSPIN_fence();       /* omitted for saccato_bug */
	start_handshake();
	wait_for_handshake();

	CAS2(from_header_fwd, from_header_copying,     FROM_OBJECT,     TRUE,	     TO_OBJECT,       FALSE,
	    success);
      }
      root = from_header_fwd; /* flip */
    }

    /* mutator */
#pragma McSPIN section
    {
      int last_written = INITIAL_VALUE;
      int readval;

      while (true) {
#pragma McSPIN nondeterministic
	{
	  ack_handshake();
	}
#pragma McSPIN nondeterministic
	{
	  last_written = 1 - last_written;
	  write(root, last_written);
	} else {
	  read(root, readval);
	  McSPIN_assert(McSPIN_variable(last_written,1,0) == McSPIN_variable(readval,1,0));
	}
#pragma McSPIN nondeterministic
	{
	  ack_handshake();
	}
      }
    }
  }
}

static inline ack_handshake()
{
  McSPIN_fence();
  if (hs_req == true)
    hs_req = false;
  McSPIN_fence();       /* removed for staccato_pso and saccato_bug */
}

static inline start_handshake()
{
  hs_req = true;
}

static inline wait_for_handshake()
{
  while (hs_req != 0)
    ;
}

/* read barrier */
static inline read(int obj, int retval) {
  if (obj == FROM_OBJECT) {
    if (from_header_fwd == FROM_OBJECT)
      retval = from_body;
    else
      retval = to_body;
  } else {
    /* assume: to_header_fwd = TO_OBJECT */
    retval = to_body;
  }
}

/* write barrier */
static inline write(int obj, int val) {
  if (obj == FROM_OBJECT) {
    if (from_header_copying == TRUE) {
      CAS2_NORET(from_header_fwd, from_header_copying,		 FROM_OBJECT,     TRUE,		 FROM_OBJECT,     FALSE);
    }
    if (from_header_fwd == FROM_OBJECT)
      from_body = val;
    else
      to_body = val;
  } else {
    /* assume: to_header_copying = FALSE */
    /* assume: to_header_fwd = TO_OBJECT */
    to_body = val;
  }
}
\end{lstlisting}

\section{McSPIN Model for Stopless}

Here we show our McSPIN model for the copying protocol of Stopless.

\begin{lstlisting}
#include "stdbool.h"
#include "atomic.h"

/* status */
#define IN_ORIGINAL 0
#define IN_WIDE     1
#define IN_COPY     2

/* value */
#define INITIAL_VALUE 0

/* space */
#define FROM_OBJECT 0
#define WIDE_OBJECT 1
#define TO_OBJECT   2

int main()
{
  /* from object */
  int from_fwd = FROM_OBJECT;
  int from_data = INITIAL_VALUE;

  /* wide object */
  int wide_status = IN_ORIGINAL;
  int wide_data;

  /* to object */
#define to_fwd TO_OBJECT
  int to_data;

  /* root */
  int root = FROM_OBJECT;
  
#pragma McSPIN parallel sections
  {
    /* collector */
#pragma McSPIN section
    {
      int x;
      int success;

      /* allocate wide object and install its forwarding pointer */
      CAS_NORET(from_fwd, FROM_OBJECT, WIDE_OBJECT);
      /* copy the payload to the wide object */
      x = wide_data;
      CAS2_NORET(wide_status, wide_data, IN_ORIGINAL, x,
                IN_WIDE, from_data);
      /* copy the payload to the final copy */
      success = 0;
      while (!success) {
	x = wide_data;
	to_data = x;
	CAS2(wide_status, wide_data, IN_WIDE, x, IN_COPY,
             x, success);
      }
      /* repoint to the final copy */
      from_fwd = TO_OBJECT;
      
      root = from_fwd;   /* flip */
    }

    /* mutator */
#pragma McSPIN section
    {
      int last_written = INITIAL_VALUE;
      int readval;
      int write_tmp;
      int write_success;
      
      while (true) {
#pragma McSPIN nondeterministic
	{
	  read(root, readval);
	  McSPIN_assert(McSPIN_variable(last_written,1,0) == McSPIN_variable(readval,1,0));
	} else {
	  last_written = 1 - last_written;
	  write(root, last_written);
	}
      }
    }
  }
}

/* read barrier */
static inline read(int obj, int retval) {
  if (obj == FROM_OBJECT) {
    if (from_fwd == FROM_OBJECT)
      retval = from_data;
    else if (from_fwd == WIDE_OBJECT) {
      if (wide_status == IN_ORIGINAL)
	retval = from_data;
      else if (wide_status == IN_WIDE)
	retval = wide_data;
      else if (wide_status == IN_COPY)
	retval = to_data;
    } else if (from_fwd == TO_OBJECT)
      retval = to_data;
  } else { /* obj == TO_OBJECT */
    /* assume: to_fwd == TO_OBJECT */
    retval = to_data;
  }
}

/* write barrier */
static inline write(int obj, int val) {
  if (obj == FROM_OBJECT) {
    if (from_fwd == FROM_OBJECT)
      CAS_NORET(from_fwd, FROM_OBJECT, WIDE_OBJECT);
    if (from_fwd == WIDE_OBJECT) {
      write_success = 0;
      while(!write_success) {
	if (wide_status == IN_ORIGINAL) {
	  write_tmp = wide_data;
	  CAS2(wide_data, wide_status, write_tmp, IN_ORIGINAL,               val, IN_WIDE, write_success);
	} else if (wide_status == IN_WIDE) {
	  write_tmp = wide_data;
	  CAS2(wide_data, wide_status, write_tmp, IN_WIDE,               val, IN_WIDE, write_success);
	} else if (wide_status == IN_COPY) {
	  to_data = val;
	  write_success = 1;
	}
      }
    } else if (from_fwd == TO_OBJECT)
      to_data = val;
  } else { /* obj == TO_OBJECT */
    /* assume: to_fwd = TO_OBJECT */
    to_data = val;
  }
}
\end{lstlisting}